\documentclass[sigconf]{acmart}

\usepackage{booktabs} % For formal tables

\usepackage{graphicx,color,colortbl,psfrag,subfig}
\usepackage{multirow}
\usepackage{flushend}
\usepackage{algorithm}
\usepackage{algpseudocode}

\usepackage{setspace}

\usepackage{amssymb}
\usepackage{amsmath}
\usepackage{amsfonts}
\usepackage{textcomp}

\setcopyright{acmcopyright}
\copyrightyear{2017} 
\acmYear{2017} 
\setcopyright{acmcopyright}
\acmConference{SIGIR'17}{}{August 7--11, 2017, Shinjuku, Tokyo, Japan}
\acmPrice{15.00}
\acmDOI{http://dx.doi.org/10.1145/3077136.3080802}
\acmISBN{978-1-4503-5022-8/17/08}

\begin{document}
\title[]{Optimizing Trade-offs Among Stakeholders in Real-Time Bidding by Incorporating Multimedia Metrics}

\author{Xiang Chen}
\affiliation{\institution{National University of Singapore}}
\email{chxiang@comp.nus.edu.sg}

\author{Bowei Chen}
\affiliation{\institution{University of Lincoln}}
\email{bchen@lincoln.ac.uk}

\author{Mohan Kankanhalli}
\affiliation{\institution{National University of Singapore}}
\email{mohan@comp.nus.edu.sg}

% ++++++++++++++++++++++++++++++++++++++++++++++++++
% ++++++++++++++++++++++++++++++++++++++++++++++++++

\begin{abstract}

Displaying banner advertisements (in short, ads) on webpages has usually been discussed as an Internet economics topic where a publisher uses auction models to sell an online user's page view to advertisers and the one with the highest bid can have her ad displayed to the user. This is also called \emph{real-time bidding} (RTB) and the ad displaying process ensures that the publisher's benefit is maximized or there is an equilibrium in ad auctions. However, the benefits of the other two stakeholders -- the advertiser and the user -- have been rarely discussed. In this paper, we propose a two-stage computational framework that selects a banner ad based on the optimized trade-offs among all stakeholders. The first stage is still auction based and the second stage re-ranks ads by considering the benefits of all stakeholders. Our metric variables are: the publisher's revenue, the advertiser's utility, the ad memorability, the ad click-through rate (CTR), the contextual relevance, and the visual saliency. To the best of our knowledge, this is the first work that optimizes trade-offs among all stakeholders in RTB by incorporating multimedia metrics. An algorithm is also proposed to determine the optimal weights of the metric variables. We use both ad auction datasets and multimedia datasets to validate the proposed framework. Our experimental results show that the publisher can significantly improve the other stakeholders' benefits by slightly reducing her revenue in the short-term. In the long run, advertisers and users will be more engaged, the increased demand of advertising and the increased supply of page views can then boost the publisher's revenue.

\end{abstract}


%\begin{CCSXML}
%<ccs2012>
%<concept>
%<concept_id>10002951.10003260.10003272.10003275</concept_id>
%<concept_desc>Information systems~Display advertising</concept_desc>
%<concept_significance>500</concept_significance>
%</concept>
%</ccs2012>
%\end{CCSXML}

%\ccsdesc[500]{Information systems~Display advertising}

%\begin{CCSXML}
%<ccs2012>
%  <concept_id>10010520.10010553.10010562</concept_id>
%  <concept_desc>Computer systems organization~Embedded systems</concept_desc>
%  <concept_significance>500</concept_significance>
%  <concept_id>10010520.10010575.10010755</concept_id>
%  <concept_desc>Computer systems organization~Redundancy</concept_desc>
%  <concept_significance>300</concept_significance>
%  <concept_id>10010520.10010553.10010554</concept_id>
%  <concept_desc>Computer systems organization~Robotics</concept_desc>
%  <concept_significance>100</concept_significance>
%  <concept_id>10003033.10003083.10003095</concept_id>
%  <concept_desc>Networks~Network reliability</concept_desc>
%  <concept_significance>100</concept_significance>
%</ccs2012>  
%\end{CCSXML}
%
%\ccsdesc[500]{Computer systems organization~Embedded systems}

% We no longer use \terms command


\maketitle

% ++++++++++++++++++++++++++++++++++++++++++++++++++
% ++++++++++++++++++++++++++++++++++++++++++++++++++

\section{Introduction}
\label{sec:introduction}

% RTB
Over the last few years, a significant development in display advertising is the emergence of RTB. It is a non-guaranteed delivery system in which page views (also called \emph{impressions}) are bought and sold via programmatic instantaneous auctions~\cite{Google_2011}. The success of RTB can be largely attributed to its user targeting and wide access. In RTB, advertisers bid for an impression from their targeted user and this impression can be from any webpage from any ad network or platform. If an advertiser wins the auction, her ad will be selected and be displayed to the user in almost real time. 

%RTB limitations
RTB has two major limitations. The first limitation is that it can not provide guaranteed delivery to advertisers. Therefore, those who want to secure and access future impressions in advance will not be satisfied~\cite{OpenX_2013}. The solution is to develop a new system that enables time-dependent allocation and pricing of future impressions. Notable examples include the recently discussed programmatic guarantee and ad options~\cite{Chen_2014_2,Chen_2015_1,Chen_2016,Wang_2012_1,Chen_2015_2}. The second limitation of RTB is that those displayed ads may not fit their hosting webpages well due to user targeting. It can be: the ad content is irrelevant to the webpage content; the color of the ad image is too light or dark to sufficiently contrast with the webpage color scheme. Ill-fitting ads will affect the effectiveness of ad branding for the advertiser, or even worse annoy users, and potentially reduce the publisher's webpage visits and revenue. It should be noted that this limitation only exists in display advertising, since sponsored search ads are displayed in terms of textual links where visual effects are not very important and also these ads are keyword-based which are well matched with the user's search query. In display advertising, this limitation is worth investigating. Firstly, it has received surprisingly little attention in the recent studies. Secondly, it is an interesting topic that lies in the area combining the research of marketing, multimedia, recommendation, and Internet economics.

\begin{figure}[t]
\centering
\includegraphics[width=0.925\linewidth]{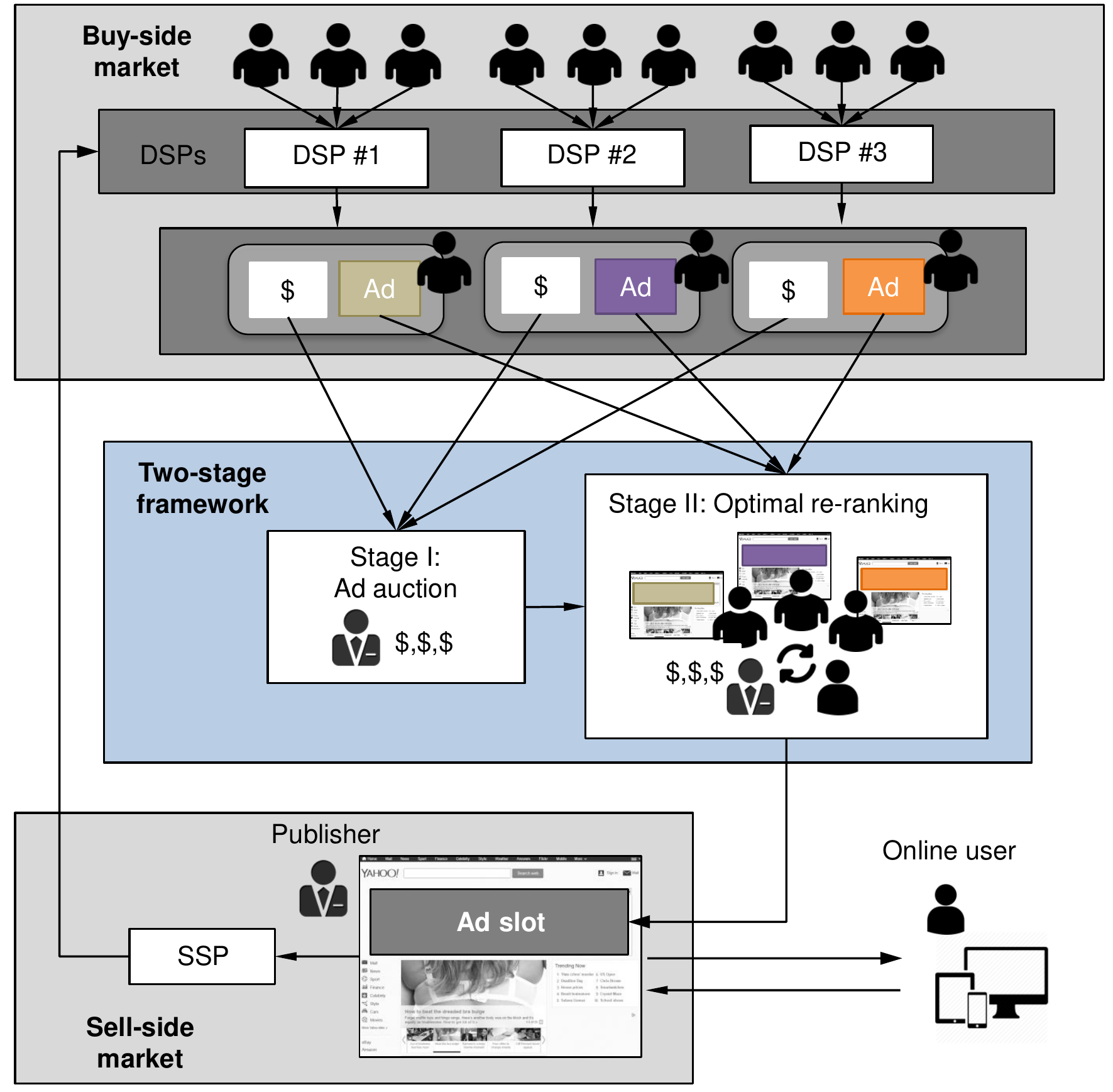}
\vspace{-7pt}
\caption{Schematic view of the two-stage framework.}
\label{fig:schematic_view}
\end{figure}

% \# what we do in this paper
In this paper, we discuss a solution to the second limitation of RTB by proposing a computational framework that optimizes trade-offs among stakeholders. As shown in Fig.~\ref{fig:schematic_view}, the framework has two stages. The first stage is based on the existing ad auction model used in RTB; and the second stage re-ranks ads based on the optimal trade-offs. In the re-ranking, there are six metric variables: the publisher's revenue, the advertiser's utility, the ad memorability, the ad CTR, the contextual relevance and the visual saliency. Revenue is always the key concern for the publisher. From an advertiser's perspective, her utility is the short-term benefit and the ad memorability is the long-term benefit. The rest three metric variables represent the user's benefits. Under the proposed framework, we discuss an algorithm that determines the optimal weights of the metric variables. In the paper, we validate the proposed framework on several datasets, including ad auction datasets and multimedia datasets. The ground truth and possible weights combinations for re-ranking are discussed. Our experimental results show that a slight decrease in the publisher's revenue can increase the performance of other variables significantly in the short term, which will also improve the engagements of both advertisers and users in the long term, and further increase the publisher's revenue.   

%\# our contributions

This paper has two major contributions. Firstly, the user's benefits have been taken into account. This is similar to the work of Bachrach et al.~\cite{Bachrach_2014}. Compared to their work, we increase the dimension of benefits by discussing trade-offs among stakeholders' multiple benefits. The framework is also tailored to RTB rather than sponsored search. Our proposed model can be regarded as a generalized framework that non-guaranteedly deliveries banner ads in RTB. Secondly, multimedia metrics have been introduced to measure some stakeholders' benefits, such as the ad memorability, the contextual relevance, and the visual saliency. These metrics naturally fit display banner ads but have not yet been discussed along with auction models in previous studies. To the best of our knowledge, this is the first work that combines multimedia metrics and auction theory together. 

The rest of the paper is organized as follows. Section~\ref{sec:related_work} reviews the related literature. Section~\ref{sec:model} discusses the proposed two-stage framework. Section~\ref{sec:experiments} presents our experimental results, and Section~\ref{sec:conclusion} concludes the paper. 

% ++++++++++++++++++++++++++++++++++++++++++++++++++
% ++++++++++++++++++++++++++++++++++++++++++++++++++
\section{Related Work}
\label{sec:related_work}

% The simplified version of the former is known as the Second-Price (SP) auction. In display advertising, the SP auction enjoys many economic properties as the VCG auction.

In online advertising, the generalized second-price (GSP) auction~\cite{Edelman_2007_2,Varian_2007} and the Vickrey-Clark-Groves (VCG) auction~\cite{Parkes_2007} have been widely used on different platforms, including display and search domains. Revenue maximization is always the key issue. Lahaie and Pennock~\cite{Lahaie_2007} proposed to use a squashing parameter to increase revenue of GSP auctions in sponsored search, and then Lahaie and McAfee~\cite{Lahaie_2011} showed that it can improve social welfare as well. Ostrovsky and Schwarz~\cite{Ostrovsky_2011} investigated the effects of optimal reserve prices for GSP auctions on Yahoo's platform. Thompson and Leyton-Brown~\cite{Thompson_2013} studied a variety of ways of increasing revenue of GSP auctions, including reserve prices and squashing parameter. Furthermore, Yuan et al.~\cite{Yuan_2014} implemented a large scale experiment and empirically compared algorithms of setting reserve price for maximizing revenue in RTB.

Trade-offs among multiple objectives or stakeholders have been discussed in several works. Likhodedov and Sandholm~\cite{Likhodedov_2003} proposed a framework that linearly combines revenue and social welfare in a single-item auction. Radlinski et al.~\cite{radlinski2008optimizing} investigated the optimization of relevance and revenue in sponsored search. Similarily, Liao et al.~\cite{liao2008adimage} combined revenue and relevance for video advertising. Lucier et al.~\cite{Lucier_2012} discussed the revenue of GSP auction at equilibrium, and demonstrated that revenue can be maximized at a non-envy-free equilibrium that also generates a socially inefficient allocation. Roberts et al.~\cite{Roberts_2013} discussed a new ranking algorithm for sponsored search auction that uses reserve prices to order the ads and discussed conditions under which revenue can be increased. Bachrach et al.~\cite{Bachrach_2014} proposed a framework that linearly combines objectives relevant to the search engine (revenue), the advertiser (welfare), and the user (clicks). %The trade-offs and constraints were discussed.   

Most related advertising research focuses on the economic benefits, little attention has been paid to the advertiser's branding effectiveness and user's experience. Therefore, we introduce several multimedia metrics in this paper. Literature in marketing and consumer psychology has already shown that the contextual relevance between the content of hosting webpage and the ads makes a large difference in their clickability~\cite{chatterjee2003modeling}, and it also has a leading effect on user's online experience~\cite{mccoy2007effects}. There exists rich research on measuring the contextual relevance using textual information. Neto et al.~\cite{ribeiro2005impedance} investigated ten strategies for a vector space model and evaluated their effectiveness individually. Although the vector space model is widely used, a more advanced matching strategy is required due to the vocabulary impedance problem among ads and webpages. Broder et al.~\cite{broder2007semantic} proposed an approach for contextual ad matching based on a combination of semantic and syntactic features. Li et al.~\cite{li2010pagesense} further measured the contextual relevance by introducing category relevance, keyword relevance, and style relevance. Category relevance measures whether the ad and webpage belong to the same category; keyword relevance measures the relevance between the keywords extracted from ads and webpage; and style relevance measures the style consistency between the ads and webpage. Zhang et al.~\cite{zhang2012advertising} proposed an advertising keywords recommendation system for short-text webpages with the help of Wikipedia. Other than textual information, images and videos are more popular on the Internet. Chen et al.~\cite{chen2010visual} allocated ads for images through image recognition techniques. To achieve better contextual relevance, Mei et al.~\cite{mei2007videosense, mei2012imagesense} and Guo et al.~\cite{guo2009adon} proposed a multimodal approach which considered both textual and visual relevance for video and image advertising. 

Although display banner ads appear on nearly every webpage, their effectiveness remains debatable. Through a series of eye-tracking experiments, recent research found that users tend to avoid ads in web search and surfing~\cite{broder2008search} and they intentionally avoid looking at such ads even when they are designed to be attention-grabbing~\cite{sajjacholapunt2014influence}. This is also known as~\emph{ad overlook}. Moreover, Owen et al.~\cite{owens2011text} explored the relationship between ads location and the degree of blindness -- the phenomenon of website users actively ignore web banner ads -- and found that users tend to ignore ads located on the bottom and right area. Intuitively, any ad that fails to capture an user's attention will be ineffective in delivering information. It has been recognized that there are five basic stages before a user accepts a message: attention, comprehension, yielding, retention and action~\cite{alwitt1985psychological}. Motivated by the above observation, in addition to the contextual relevance, we further introduce two other factors to increase user engagement towards banner ads: the visual saliency and the ad image memorability. The former increases the probability that an user will notice the displayed ad while the latter improves an user's brand perception after she sees the ad image.

%Online advertising is a multidiscipline field which involves economics (auction, RTB, guarantee display), multimedia (contextual relevance, user experience), and psychology (user behaviour). However, most existing works try to improve the effectiveness of online advertising only for one specific aspect. Since it is an interplay among publisher, advertiser and user~\cite{Broder_2008,YYuan_2014}, we believe that by considering the utilities or interests of all players, the whole system will be optimized on the long run. 

\begin{table}[t]
\small
\centering
\caption{Notations.}
\label{tab:notations}
\vspace{-7pt}
\begin{tabular}{r|p{2.75in}}
\hline
\hspace{-20pt} Notation \hspace{-5pt} & Description\\
\hline
$Z, \widetilde{Z}$  & Set of auctions\\
$N_z$   & Set of advertisers in auction $z, z \in Z$\\
$K$     & Set of metric variables for Stage II re-ranking, $|K| = 6$\\
$b_{i,z}$ & Bid price of advertiser $i$ in auction $z$, $i \in N_z, z \in Z$\\
$b_{(i),z}$ & $i$th highest bid in auction $z$, $i \in N_z, z \in Z$\\
$y_{i,j,z}$ & Probability that advertiser $i$ is allocated to slot $j$ in auction $z$\\
$p_{i,z}(b_z)$ & Payment of advertiser $i$ in auction $z$, $b_z=(b_{1,z}, \cdots, b_{n_z,z})$\\
$\omega_k$ & Weight of metric variable $k$ in the re-ranking, $k \in K$\\
$x_{k,*,z}$ & Value of metric variable $k$ for the selected advertiser in auction $z$ by our proposed model\\
$x_{k,\neg,z}$ & Value of metric variable $k$ for the selected advertiser in auction $z$ by the SP auction (the ground truth)\\
$\xi_{k,Z}$ & Average change of metric variable $k$ of auctions in set $Z$\\
$\theta_k$ & Threshold for the change of metric variable $k$\\
$rs_{i,z}$ & Rank score of advertiser $i$ in auction $z$\\
$rs_{*,z}$ & Rank score of the selected advertiser in auction $z$ by our proposed model\\
$rs_{\neg,z}$ & Rank score of the selected advertiser in auction $z$ by the SP auction (the ground truth)\\
\hline
\end{tabular}
\end{table}

% ++++++++++++++++++++++++++++++++++++++++++++++++++
% ++++++++++++++++++++++++++++++++++++++++++++++++++
\section{The Two-Stage Framework}
\label{sec:model}

The proposed framework consists of two stages. The first stage is based on the second-price (SP) auction model. The second stage re-ranks ads based on the weighted linear combination of benefits of multiple stakeholders, which ensures the selected ad will optimize the trade-offs among them. For the reader's convenience, the key notations are summarized in Table~\ref{tab:notations}. 

\subsection{Stage I: Ad Auction}
\label{sec:stage_1}

In RTB, the SP auction model has been widely used. Consider auction $z \in Z$, a single impression is bidded by $n_z$ advertisers, their cost-per-impression\footnote{In display advertising industry, the cost-per-impression is usually quoted in terms of cost-per-mille (CPM) which represents the price of 1000 impressions.} bids are denoted by a bid vector $b_z = (b_{1,z}, \cdots, b_{n_z,z})$, and their true values are denoted by a value vector $v_z = (v_{1,z}, \cdots, v_{n_z,z})$. In the simplest scenario, the advertiser with the highest bid will be selected and her ad will be displayed to the user. When the ad is displayed, the advertiser then pays the publisher for this page view. In order to re-rank ads in the second stage, we need to know what would be the possible payments from other advertisers. To achieve this, $n_z-1$ pseudo ad slots are created so that the payments from other advertisers can be estimated.  

We denote the auction outcome by $\{y_{i,j,z}, p_{i,z}\}_{i \in N_z, j \in N_z, z \in Z}$, where $y_{i,j,z}$ is the probability that advertiser $i$ is allocated to slot $j$ in auction $z$, and $p_{i,z}$ is her payment. Note that slots are ranked so that slot $j \geq 2$ is the created pseudo slot. The auction outcome satisfies the following conditions: 
\begin{align}
0 \leq y_{i,j,z} \leq 1, \\
\sum_{i \in N_z} y_{i,j,z} \leq 1,\\
\sum_{j \in N_z} y_{i,j,z} \leq 1.
\end{align}

Let $b_{(i),z}$ be the $i$th highest bid in $b_z$, then 
\begin{equation}
y_{i,j,z}(b_z) =  \mathbb{I}_{\{b_{i,z} = b_{(j),z}\}}, 
\end{equation}
where $\mathbb{I}_{\{\cdot\}}$ is the indicator function. Then advertiser $i$'s payment is
\begin{equation}
p_{i,z}(b_z) = \sum_{j \in N_z} b_{(j+1),z} y_{i,j,z}(b_z), 
\end{equation}
where $b_{(n_z+1), z}$ is equal to the reserve price, whose value can be set to be any value between $0$ and $b_{(n_z), z}$.

\begin{table*}[tp]
\small
\centering
\caption{Specifications of metric variables for Stage II.}
\label{tab:second_stage_var_spec}	
\vspace{-7pt}
\begin{tabular}{l|l|l|l|l}
\hline
\multirow{2}{*}{Variable}   & \multicolumn{3}{c|}{Computation} & \multirow{2}{*}{Input for Stage II}\\
\cline{2-4}
           & Method & Input & Output         & \\
\hline
Publisher's revenue    & Stage I      & Bid & Payment & $x_{1,i,z} = $ normalized $p_{i,z}$\\
Advertiser's utility    & Stage I      & Bid, payment & Utility    & $x_{2,i,z} = $ normalized $(v_{i,z} - p_{i,z})$\\
Ad memorability & MemNet~\cite{khosla2015understanding} & Ad image & Memorability score & $x_{3,i,z} = $ normalized score\\
Ad CTR    & Given by data         & CTR &  CTR & $x_{4,i,z} = $ normalized CTR\\
Contextual relevance & TakeLab~\cite{vsaric2012takelab} & Title, keywords, description & Semantic similarity score & $x_{5,i,z} = $ normalized score\\
Ad saliency & MBS~\cite{zhang2015minimum} & Web page snapshot, ad image & Saliency map and ratio & $x_{6,i,z} = $ normalized ratio \\
\hline 
\end{tabular}
\end{table*}

\subsection{Stage II: Optimal Re-Ranking}
\label{sec:re-ranking}

All advertisers are re-ranked in the second stage. The rank score of advertiser $i, i \in N_z$, is defined as follows:
\begin{align}
rs_{i,z} = & \ \sum_{k \in K} \omega_k x_{k, i, z},
\end{align}
where $x_{k, i, z}$ is the input value of metric variable $k$, $\omega_k$ is its weight, and $k \in K = \{1,\cdots,6\}$. Below we discuss what are these
 variables, how to obtain their input values and their weights in the re-ranking.

\subsubsection{Specifications of Metric Variables}

Table~\ref{tab:second_stage_var_spec} specifies our metric variables in the re-ranking, including the publisher's revenue, the advertiser' utility, the ad memorability, the ad CTR, the contextual relevance, and the ad saliency.  The input values of these metric variables in the second stage are denoted by $x_{k,i,z}, \forall k \in K$, respectively. They are the normalized values lying between 0 and 1 by using the min-max method for a set of auctions.

Revenue is always the key issue for the publisher. In rank score $rs_{i,z}$, it can be measured by advertiser $i$'s (normalized) payment, whose value can be obtained from ad auction in Stage I. 

For advertiser $i$, her benefit can be measured at short-term and long-term levels. The short-term benefit is her utility, which is defined as the difference between her value and payment. It shows the advertiser's cost saving. In the long run, the ad's memorability is an important metric, particularly, for branding purpose~\cite{Neumeier_2005}. It shows how likely the user will remember the advertiser's ad a few weeks or months later. Here we employ the MemNet model~\cite{khosla2015understanding} to predict visual memorability of ad image. It is a convolutional neural network (CNN) trained in an annotated image memorability dataset, where the input is an ad image and the output is a single real value which indicates the memorability of this image. The higher the value is, the more memorable is the image.

The user's benefit can be measured by her page view experience, including the ad CTR, the contextual relevance and the ad saliency~\cite{Lehmann:2012:MUE:2358968.2358983}. The ad CTR is defined as the number of clicks on the advertiser's ad divided by the number of displays, whose value is usually given by data or can be estimated from historical advertising records. In essence, the ad CTR is an ad quality metric -- a high CTR means that the advertiser's ad is attractive and also more relevant to the user's needs. The contextual relevance measures if the ad content is more or less relevant to its hosting webpage content. If an ad is more relevant, it will be less intrusiveness~\cite{mei2012imagesense}. In the paper, we use the TakeLab system~\cite{vsaric2012takelab} to measure the semantic similarity of textual contents between the ad and the webpage. TakeLab uses a support vector regression model with multiple features measuring the word-overlap similarity and syntax similarity. The input is two textual sentences and the output is the semantic similarity score. The score ranges from 0 to 5, where 0 indicates irrelevant and 5 indicates totally relevant. We construct the textual information of the webpage and the ad by using the webpage title, keywords and description, etc. An advantage of TakeLab is that it has an outstanding performance in measuring the similarity between short text snippets, while most other algorithms focus on large documents. The ad saliency metric measures whether the ad image can be easily spotted within its hosting webpage. The more salient ads tend to draw more user's attention~\cite{xiang2015salad}. In the paper, we use the minimum barrier salient (MBS) object detection method~\cite{zhang2015minimum} to calculate the saliency of the ad image. The input is image screenshots of ad and webpage, and the output is the corresponding saliency map. For each pair of webpage and ad candidate, we embed the ad into the webpage and use the MBS method to compute the saliency map and the saliency score is the mean value of each pixel within the ad area. 

\subsubsection{Determination of the Optimal Weights}
\label{sec:optimal_weights}

The weights of variables in the rank score determine the optimal trade-offs. As described earlier, the publisher needs to sacrifice a certain amount of revenue in the short term in order to increase the benefits of other stakeholders. Let $\theta_{1}$ be the maximum pre-determined loss rate of revenue and let $\theta_{k},\forall k \in K\backslash\{1\}$, be the minimum increase rate of other variables. Given the maximum loss and the minimum increase targets, we can obtain the optimal weights from the training set $\widetilde{Z}$. 

Our algorithm can be expressed as follows:   
\begin{align}
\max_{\omega_1, \cdots,\omega_6} 
& \ \sum_{z \in \widetilde{Z}} rs_{*,z}, \\[0.02in]
\text{subject to} 
& \ 0 \leq \omega_{k} \leq 1, \forall k \in K, \label{eq:weight_nonnegative}\\[0.02in]
& \ \sum_{k \in K} \omega_{k} = 1, \label{eq:weight_unity}\\[0.05in]
& \ |\xi_{1, \widetilde{Z}}| \leq |\theta_{1}|, \theta_1 \leq 0, \label{eq:threshold_revenue} \\[0.02in]
& \ \xi_{k, \widetilde{Z}} \geq \theta_{k}, \theta_k \geq 0, \forall k \in K\backslash\{1\}, \label{eq:threshold_others} 
\end{align}
where $rs_{*,z}$ is the rank score of the selected advertiser in auction $z$ by our proposed model, $\xi_{k, \widetilde{Z}}$ is the mean of changes of variable $k$ between our proposed model and the ground truth, defined by
\begin{equation}
\xi_{k, \widetilde{Z}} = \frac{\sum_{z\in \widetilde{Z}}(x_{k, *, z}-x_{k,\neg,z})}{\sum_{z \in \widetilde{Z}}x_{k,\neg, z}}, \label{eq:changes_of_variables}
\end{equation}
where $x_{k,*,z}$ is the input value of metric variable $k$ for the selected advertiser in auction $z$ by our proposed model, $x_{k,\neg,z}$ is the input value of metric variable $k$ for the selected advertiser in auction $z$ in the ground truth. 

The optimal weights maximize the sum of rank scores of the select advertisers from all auctions in the training set. Eqs.~(\ref{eq:weight_nonnegative})-(\ref{eq:weight_unity}) ensure each variable has an impact in the re-ranking but its impact has an upper bound. Eqs.~(\ref{eq:threshold_revenue})-(\ref{eq:threshold_others}) further specify the lower bounds of trade-offs where the maximum decrease of the publisher's revenue and the minimum increases for other variables. 

%, and $\theta_k$ is the threshold value for specifying the bound of changes in variable $k$.

% (represented by $\theta_{k},\forall k \in K\backslash\{1\}$)
% The optimal weights vary under different $\theta_{k}$. Below we discuss an algorithm that can obtain the optimal weights from a given training set $\widetilde{Z}$. 

\begin{table}[t]
\small
\centering
\caption{Multimedia datasets.}
\vspace{-7pt}
\label{tab:multimedia_datasets}
\begin{tabular}{r|r|r}
\hline
Dataset & AOL & YouTube\\
\hline
Ad network & AOL ONE  & Google AdSense\\
From & 06 Sept 2016 & 06 Sept 2016 \\
To   & 09 Sept 2016 & 09 Sept 2016 \\
Location & Singapore & Singapore \\
\# of unique webpages & 5,243 & 7,173\\
\# of unique ads & 96 & 539 \\
\hline
\end{tabular}
\vspace{10pt}
\caption{Ad auction datasets.}
\label{tab:auc_datasets}
\vspace{-7pt}
\begin{tabular}{r|r|r}
\hline
Dataset & SSP & Microsoft\\ 
\hline
Ad type    & Display  & Search\\
Ad auction & SP (RTB) & GSP   \\
Market  & UK & US \\
From    & 08 Jan 2013 & 26 Dec 2011\\
To      & 14 Feb 2013 & 03 Mar 2012\\
\# of ad slots    & 31 & 4,376      \\
\# of user tags   & NA & NA        \\
\# of publishers & NA & 1 \\
\# of advertisers   & 374      & NA\\ 
\# of auctions      & 6,646,643  & 35,550\\ 
\# of bids          & 33,043,127 & NA\\
Bid quote               & GBP/CPM  & GBP/CPC\\
Bids of each auction & $\surd$ & NA\\
Winning bid          & $\surd$ & $\surd$\\
Winning payment      & $\surd$ & $\surd$\\
Estimated CTR        & NA      & $\surd$\\
\hline
\end{tabular}
\vspace{10pt}
\caption{Summary of CTRs in the Microsoft dataset.}
\label{tab:microsoft_dataset}
\vspace{-7pt}
\begin{tabular}{r|r|r|r|r}
\hline
Position & Minimum & Maximum & Mean    & Std.   \\
		 & of CTR  & of CTR  & of CTR  & of CTR \\
\hline
1  &     0.00\textperthousand  & 333.30\textperthousand  &   13.97\textperthousand & 28.49\textperthousand  \\
2  &     0.00\textperthousand  & 200.00\textperthousand  &   7.97\textperthousand  & 15.30\textperthousand  \\
3  &     0.00\textperthousand  & 750.00\textperthousand  &   5.66\textperthousand  &  12.47\textperthousand \\
4  &     0.00\textperthousand  & 90.90\textperthousand   &   4.53\textperthousand  &  9.68\textperthousand  \\
5  &     0.00\textperthousand  & 76.90\textperthousand   &   1.53\textperthousand  &  3.58\textperthousand  \\
6  &     0.00\textperthousand  & 333.30\textperthousand  &   1.11\textperthousand  &  3.62\textperthousand  \\
7  &     0.00\textperthousand  & 1000.00\textperthousand &   0.68\textperthousand  &  5.73\textperthousand  \\
8  &     0.00\textperthousand  & 108.10\textperthousand  &   0.51\textperthousand  &  1.65\textperthousand  \\
\hline
\end{tabular}
\end{table}

\subsection{Discussion}
\label{sec:discussion}

Display advertising is an interplay among different stakeholders. A vigorous and healthy advertising eco-system should consider the benefits of all stakeholders. Compared with the existing advertising systems that select ads mainly from the publisher's interest, our proposed framework select ads based on the optimized trade-offs among stakeholders. The incorporated multimedia metrics are computed by the state-of-the-art multimedia techniques, including the MemNet model for the ad image memorability~\cite{khosla2015understanding}, the TakeLab method for the semantic similarity~\cite{sajjacholapunt2014influence}, and the MBS method for the visual saliency~\cite{zhang2015minimum}. Moreover, our framework can be extended with more metrics, such as ad interestingness and image aesthetic. 

It should be also noted that the proposed two-stage framework does not ensure advertisers to be truth-telling in Stage I. If an advertiser is truth-telling, her utility can be obtained by computing the difference between her bid and payment. If she is not truth-telling, her value can be learnt from historical ad auctions in Stage I. Value estimation models were discussed by~Athey and Nekipelov~\cite{Athey_2010}, and Pin and Key~\citep{Pin_2011}, respectively. Similarly, we can assume each advertiser is optimal -- who maximizes her expected utility in the bidding -- and then estimate her value from the training data, and the data should be obtained from the two-stage framework rather than the existing RTB campaigns.  On the other hand, we think the proposed framework will not significantly affect an advertiser's truthfulness. Firstly, the three multimedia metric variables and the ad CTR are exogenous. They do not directly affect an advertiser's bidding behaviour. Secondly, advertisers' payments are computed by creating pseudo slots. If an advertiser is not the winning advertiser in the ground truth but she is selected after re-ranking, her actual payment in the two-stage framework is equal to the payment from the SP auction without other advertisers who have higher bids than her. Since advertisers' bids are not disclosed, compared to the existing RTB, the winning advertiser has no further information to support her nontruth-telling biddings. 

%As they are truthful-telling under the SP auction~\cite{Parkes_2007}, their bids reveal their values, so $v_z = b_z$. 

% ++++++++++++++++++++++++++++++++++++++++++++++++++
% ++++++++++++++++++++++++++++++++++++++++++++++++++
\begin{figure}[tp]
\centering
\includegraphics[width=0.95\linewidth]{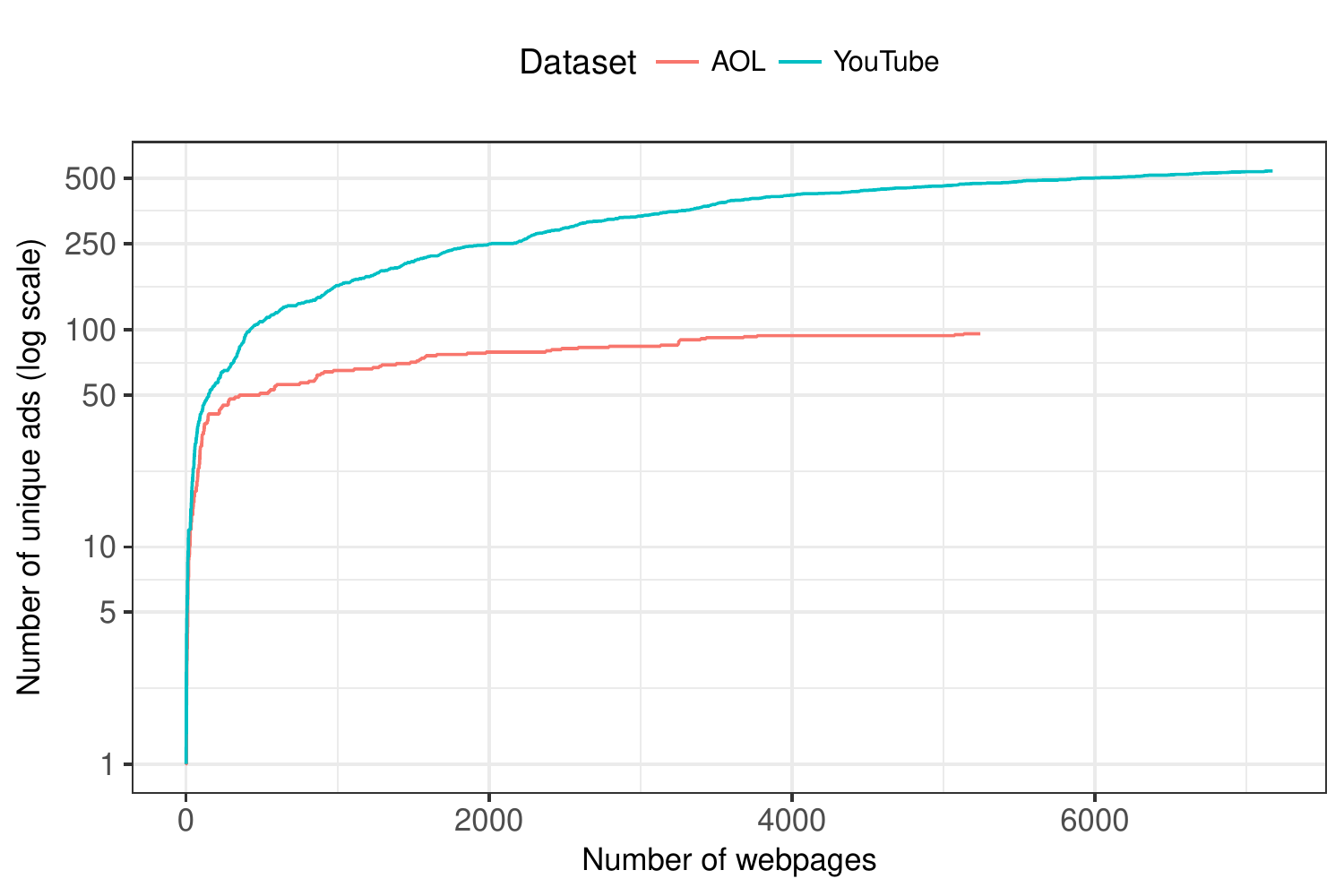}
\vspace{-7pt}
\caption{Unique ads in the multimedia datasets.}
\label{fig:unique_ad_with_webpage}
\end{figure}

\section{Experiments}
\label{sec:experiments}

This section describes our datasets, experimental settings, variable analysis, sensitivity analysis, and overall results. 

\begin{figure*}[t]
\centering
\includegraphics[width=0.975\linewidth]{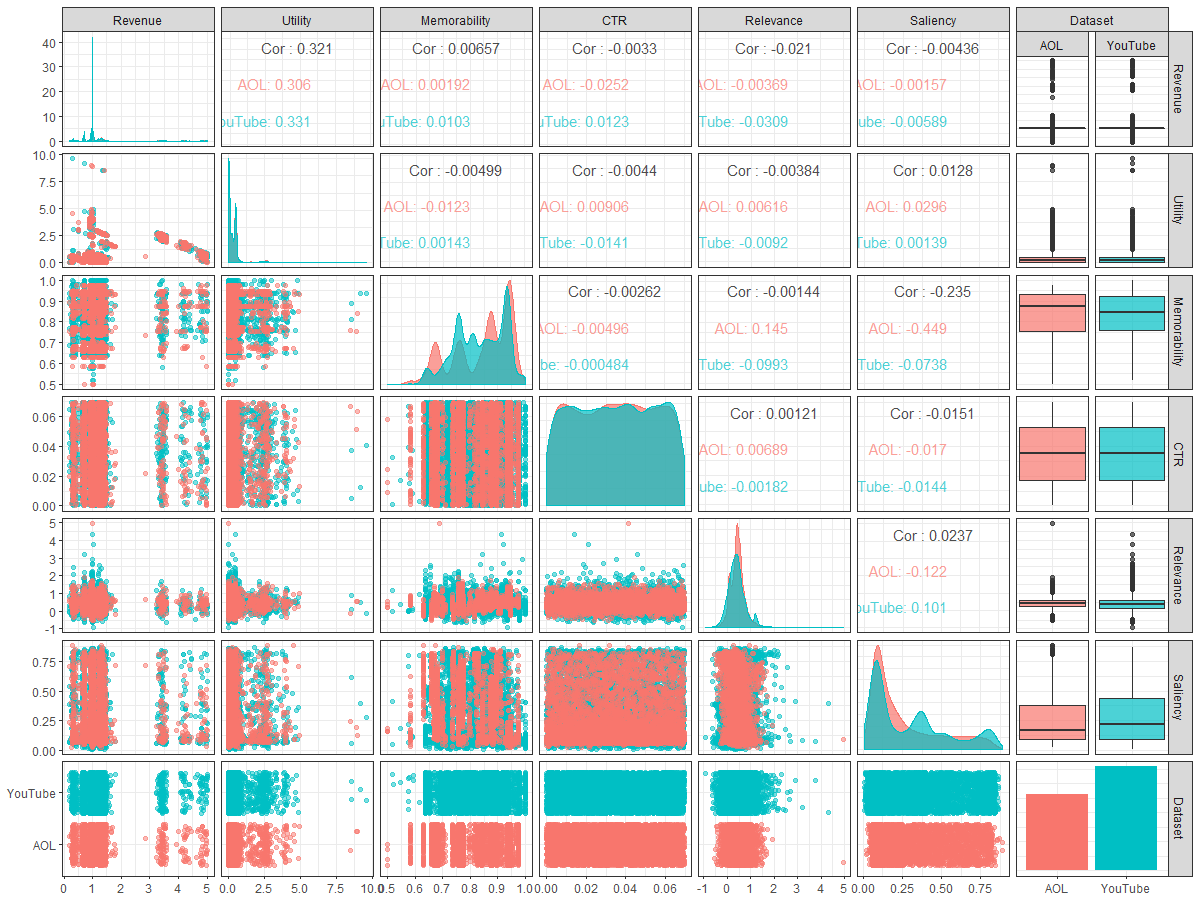}
%\vspace{-10pt}
\caption{Relationship among variables in the ground truth data (before normalization).}
\label{fig:variable_relationship}
\end{figure*}

\subsection{Datasets}
\label{sec:data_description}

Four distinct datasets are used in our experiments, including two multimedia datasets and two ad auction datasets. The two multimedia datasets contain the data collected from AOL and YouTube over the period from 06 September to 09 September in 2016. AOL is an article sharing website while YouTube is a video sharing website. The two platforms use different ad networks: AOL uses AOL Advertising; YouTube uses Google AdSense. Both datasets were collected in Singapore. As we focus on the single slot ad displaying, each webpage in our datasets contains only one ad slot. We start collecting the data from a particular seed URL that contains multiple categories of content.\footnote{
\small
\begin{tabular}{ll}
AOL: & \href{http://www.aol.com}{\tt\small http://www.aol.com}\\
YouTube: & \href{https://www.youtube.com/channels}{\tt\small https://www.youtube.com/channels}
\end{tabular}
} In this way, the diversity of webpages and banner ads can be ensured. In collecting the data, the web browser is set in the privacy mode, which disables browsing history, web cache and data storage in cookies so that the collected banner ads are not affected by the previous page views. In each dataset, we extract the ads from their webpages to create a set of banner ads and a set of webpages with blank ad slots. For each webpage, the collected data includes title, keywords, description, whole webpage snapshot, ad image. We also crawl title, keywords and description from the ad landing page (i.e., the delivered webpage if an ad is clicked by the visiting user). Note that we do not consider animation ads in our banner ads. When collecting text information about the webpage for YouTube, we also crawl video tags. All the collected text data can be viewed as a summary about the webpage, which can be used to measure the contextual relevance with ad candidates in experiments. Table~\ref{tab:multimedia_datasets} describes our multimedia datasets. We collect 5,243 unique webpages from AOL and 7,173 unique webpages from YouTube, and the number of their corresponding unique banner ads are 96 and 539, respectively. Fig.~\ref{fig:unique_ad_with_webpage} shows the growth trends of unique ads on both platforms. Although there are a large number of impressions in the ad network, only a few ads are displayed. We find that some ads re-appear from time to time. The repetitive display strategy reinforces users' memory for branding but it also is a source of intrusiveness into users' online experience~\cite{campbell2008shut}. 

The two ad auction datasets contain campaign information for display advertising and sponsored search. Table~\ref{tab:auc_datasets} provides a brief summary of both datasets. For display advertising, the dataset reports 33,043,127 bids in 6,646,643 auctions over 31 ad slots from a medium-size supply-side platform (SSP) in the UK over the period from 08 January 2013 to 14 February 2013. This SSP dataset has also been used in several recent online advertising studies~\cite{Chen_2014_2,Chen_2016,Yuan_2014} and is used as the major dataset in our experiments. In the other ad auction dataset, the estimated campaign results of 547 keywords were crawled from Microsoft adCentre (now Bing Ads) over the period from 26 Dec 2011 to 03 Mar 2012. In adCentre, the targeted keywords, budget, and other settings like matching types were submitted to the system. It then returned a list of estimated statistics, including the estimated cost-per-click (CPC), clicks, and impressions based on 8 positions, which are ranked from top to bottom in the mainline paid listing and sidebar paid listing of search engine result pages. The Microsoft dataset provides information of CTRs, which is important in our proposed model but is not included in the SSP dataset. Table~\ref{tab:microsoft_dataset} summarizes CTRs across keywords in the Microsoft dataset. Since we focus on the single slot ad displaying, we use the CTR estimated from the first ad position from the mainline paid listing to simulate the CTR in our experiments. %Simply, the CTR of each ad is generated from a uniform distribution in the range of 0 and 13.97\textperthousand $+$ 1.96 $\times$ 28.49\textperthousand.  

%the auction dataset as used in \cite{chen2014dynamic}, and two multimedia datasets including webpage dataset and ad dataset, collected from two popular social platforms in Singapore. In the SSP auction dataset, each record only provide a series of bidding prices. Note that, all the bids are expressed as Cost-Per-Mille (CPM). We cannot get any information about the ad slots, ads and advertisers. In this regard, we associate the bidding price with an ad through random sampling. In DSP auction dataset, we can figure out the CTR of each ads, which proves evidence of simulating CTR for each ad in our ad-dataset. More details can be found in section \ref{sec:experimental_setting}.

% ++++++++++++++++++++++++++++++++++++++++++++++++++
\vspace{-5pt}
\subsection{Experimental Settings}
\label{sec:experimental_setting}
\begin{figure*}[t]
\centering
\begin{minipage}{1\linewidth}
\small
\centering
\begin{tabular}{ccc}
Webpage image & Ad images & Variables for re-ranking (after normalization)\\
	\begin{minipage}{0.25\linewidth}
	\includegraphics[width=1\textwidth]{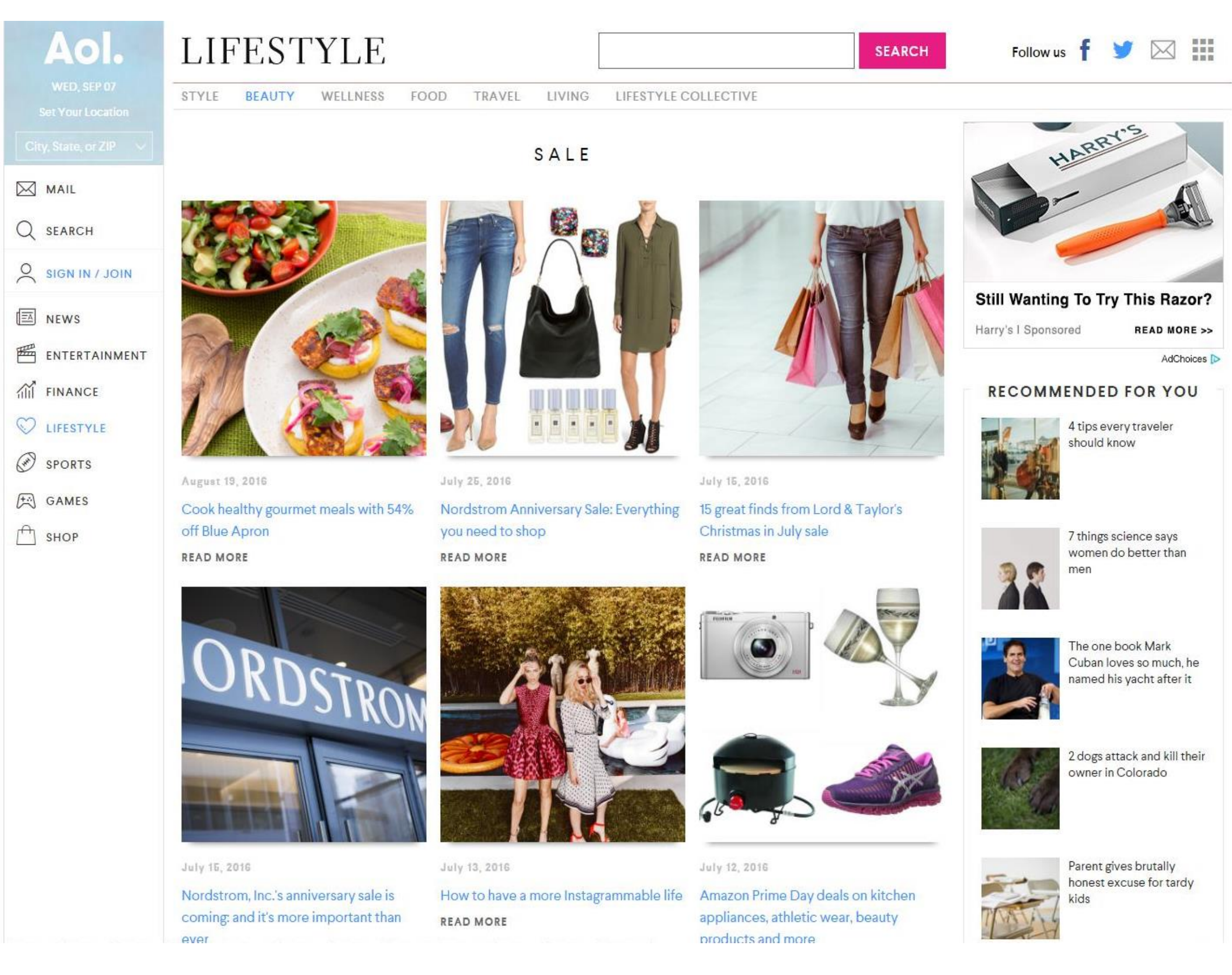}
	\end{minipage}
	& 
	\begin{minipage}{0.22\linewidth}
	\vspace{5pt}
	\includegraphics[width=1\textwidth]{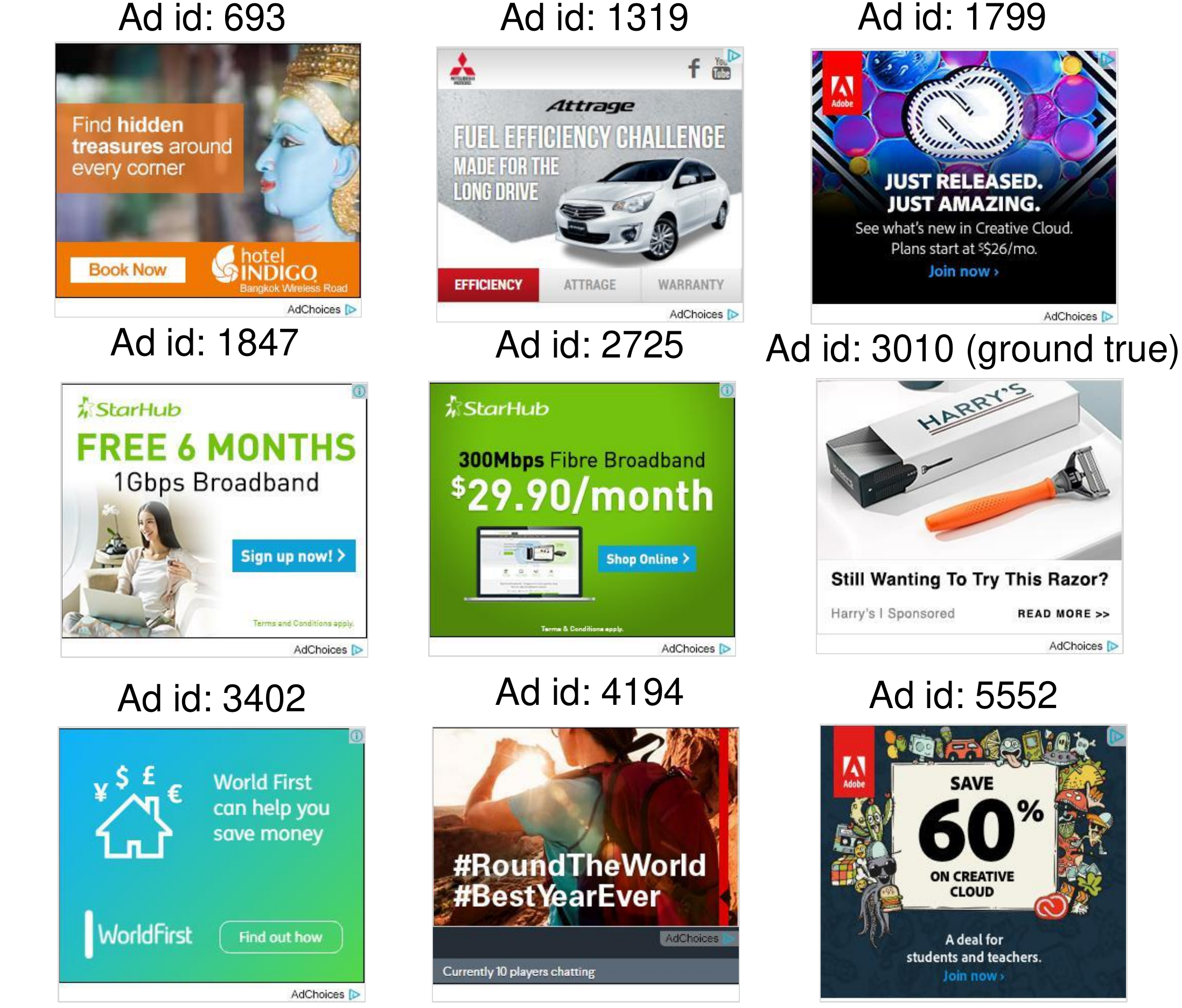}
	\end{minipage}
	&
	\small
	\hspace{5pt}
	\begin{tabular}{  r | r | r | r | r | r | r }
	\hline
	Ad id & $x_{1}$ & $x_{2}$ & $x_{3}$ & $x_{4}$ & $x_{5}$ & $x_{6}$ \\
		\hline
		693		&   0.1999	&	0.0000	&	0.7164	&	0.9387	&	0.1699	&	0.7286\\
		1319	&	0.0400	&	0.0000	&	0.8277	&	0.4077	&	0.2187	&	0.1639\\
		1799	&	0.0160	&	0.0264	&	0.5567	&	0.3353	&	0.3698	&	0.8360\\
		%1847	&	2.00e-05	&	0.0176	&	0.8971	&	0.3698	&	0.2671	&	0.1025\\
		1847	&	0.0000	&	0.0176	&	0.8971	&	0.3698	&	0.2671	&	0.1025\\
		%2725	&	2.00e-05	&	0.0000	&	0.9244	&	0.0712	&	0.2617	&	0.8763\\
		2725	&	0.0000	&	0.0000	&	0.9244	&	0.0712	&	0.2617	&	0.8763\\
		\textbf{3010}	&	\textbf{0.1999}	&	\textbf{0.1101}	&	\textbf{0.9139}	&	\textbf{0.2596}	&	\textbf{0.2734}	&	\textbf{0.1059}\\
		3402	&	0.1441	&	0.0614	&	0.8950	&	0.7269	&	0.2361	&	0.7804\\
		4194	&	0.0400	&	0.0000	&	1.0000	&	0.0720	&	0.2163	&	0.2629\\
		5552	&	0.0400	&	0.1148	&	0.5420	&	0.2836	&	0.3405	&	0.8823\\
		\hline
	\end{tabular}\vspace{5pt}
\end{tabular}\\[0.1in]
(a)\\
\vspace{5pt}
\end{minipage}
\begin{minipage}{0.475\linewidth}
\small
\centering
\includegraphics[width=0.975\textwidth]{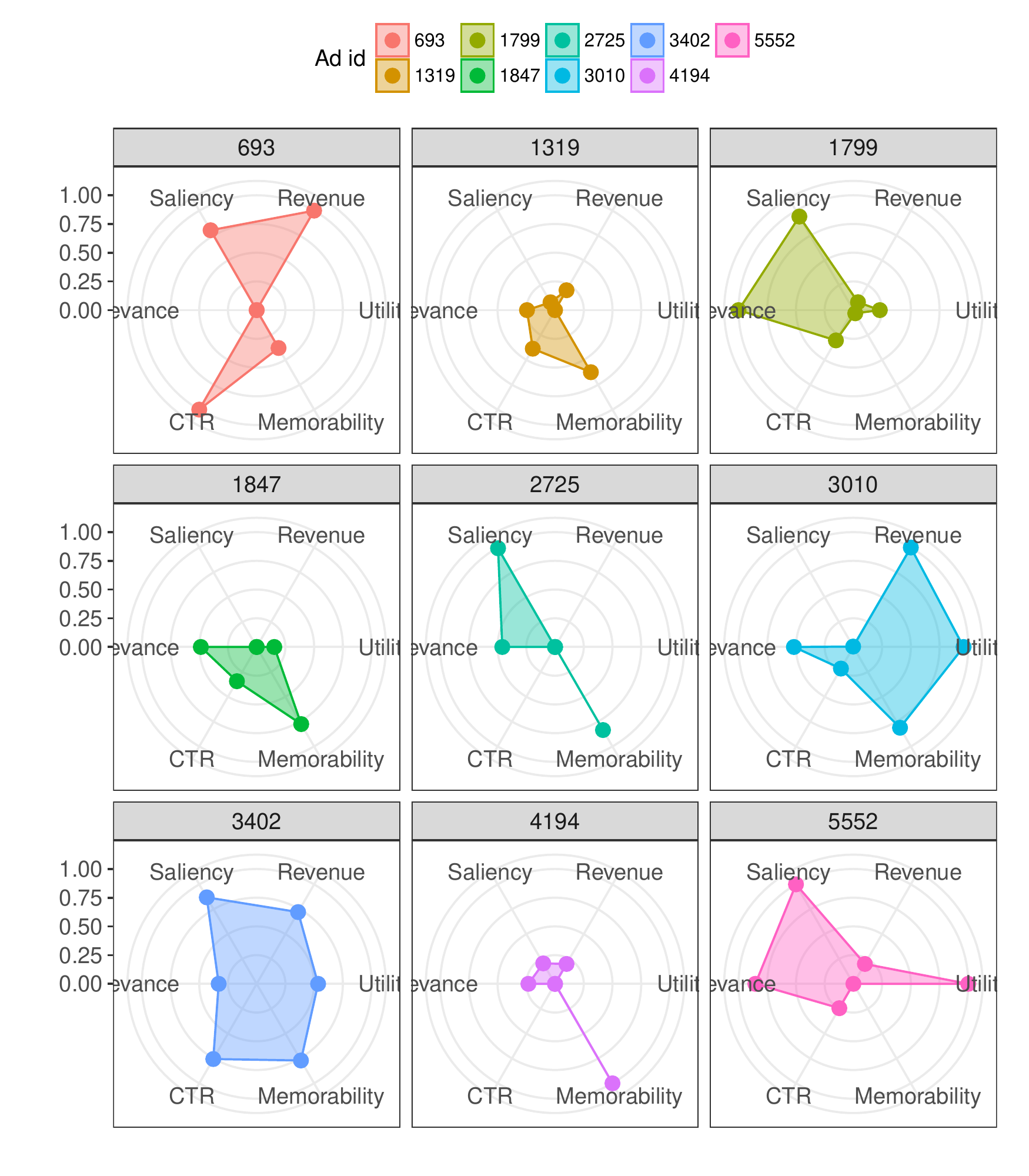}\\\vspace{5pt}
(b)
\\
\vspace{5pt}
\end{minipage}
\hspace{15pt}
\begin{minipage}{0.475\linewidth}
\small
\centering
\includegraphics[width=0.975\textwidth]{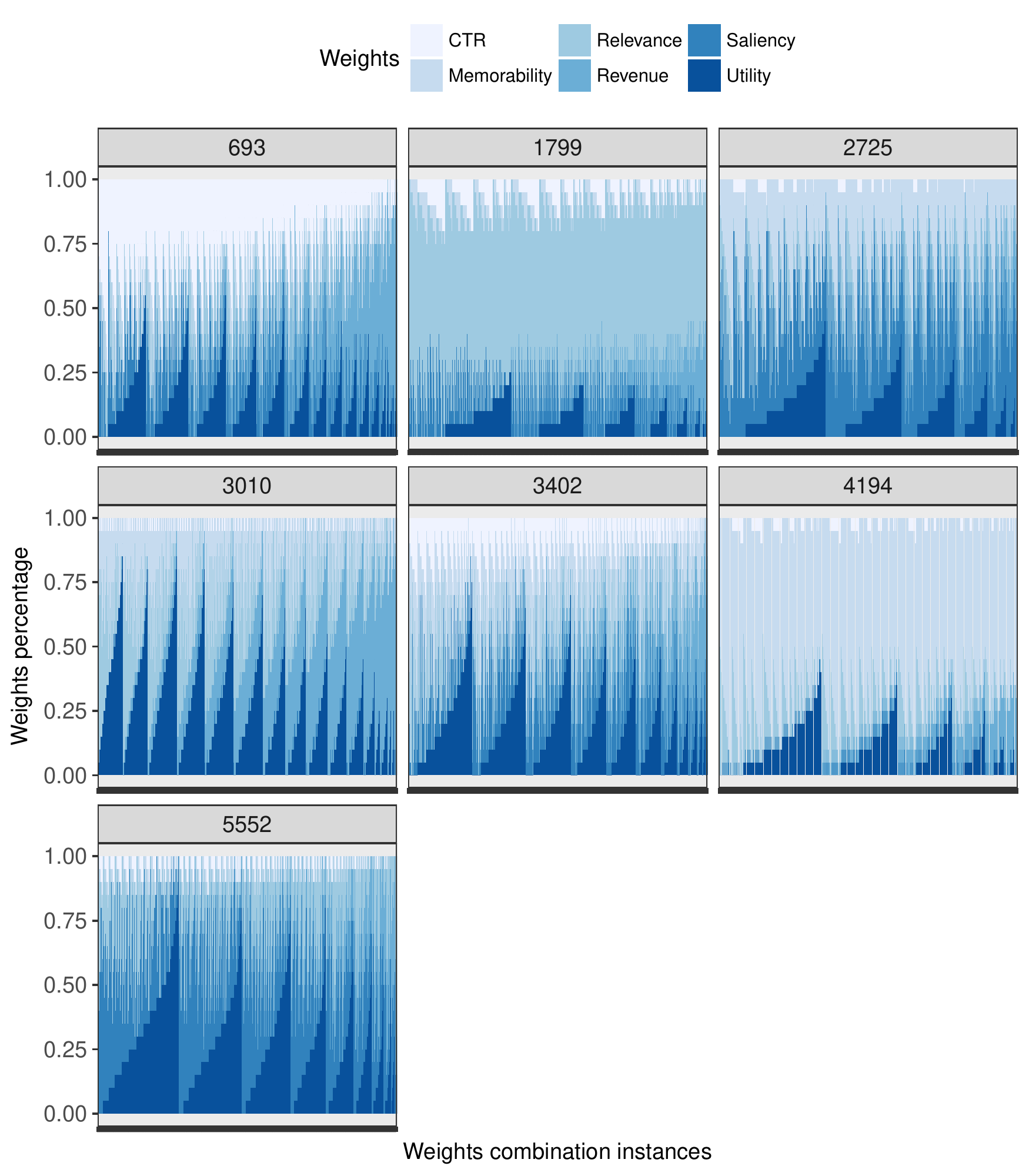}\\\vspace{5pt}
(c)\\
\vspace{5pt}
\end{minipage}
\vspace{-5pt}
\caption{Empirical example of sensitivity analysis of re-ranking for an AOL page view: (a) the raw data in the auction where the ground truth has been highlighted in bold; (b) the comparison of the ground truth ad with other candidates; (c) sensitive analysis of weights in this auction. }
\label{fig:sensitivity_analysis}
\end{figure*}

As described in Section~\ref{sec:data_description}, each of our datasets only provides partial information: the SSP dataset provides advertisers' bids for each auction which can also be used for computing the revenue and the utility; the Microsoft dataset provides CTR distributions on different ad positions; the AOL and YouTube datasets contain screenshots of ads and webpages and their textual descriptions, which can be used to calculate the multimedia metrics such as ad image memorability, the contextual relevance, and the visual saliency. To simulate a real world display advertising environment, we randomly sample and match records and metrics to connect our auction datasets with the multimedia datasets so that each ad will have both bidding campaign records and multimedia metrics. Given a webpage from the multimedia datasets, we use the following five steps to reproduce its corresponding auction:
\begin{algorithmic}[0]
\State \textbf{Step 1: Sampling an auction in RTB}\\ 
We randomly sample an auction from the SSP dataset. The auction contains a set of advertisers and their bids. 

\State \textbf{Step 2: Creating the ground truth data}\\
Since the given webpage has an ad displayed when it is web crawled. This original ad is considered as the ground truth. 

\State \textbf{Step 3: Sampling candidate ads}\\ 
Let $n$ denote the number of advertisers in the sampled auction in Step 1. We then randomly sample $n-1$ ads from the same multimedia dataset of the webpage. To ensure the uniqueness of each ad in the auction, we exclude the original ad when we sample $n-1$ ads from the multimedia dataset. 

\State \textbf{Step 4: Matching ads and bids}\\ We match the $n$ bids with $n$ ads. Since the original ad is displayed, we consider it belongs to the winning advertiser in the auction and match it with the highest bid. The rest $n-1$ ads and $n-1$ bids are then randomly matched. 

\State \textbf{Step 5: Simulating CTRs}\\ For each ad, we generate its CTR from a uniform distribution in the range of 0 and 13.97\textperthousand $+$ 1.96 $\times$ 28.49\textperthousand. This is based on the mean and std. of CTRs in the first position in the Microsoft dataset, as also shown in Table~\ref{tab:microsoft_dataset}.

\end{algorithmic}

%%\noindent\textbf{Step 1:} We first randomly sample an auction from our SSP auction dataset, which consists of $n$ bidding price.\\
%%\noindent\textbf{Step 2:} We then randomly sampling $n-1$ ads from our ad dataset. Note that we only sample $n-1$ ads because there already exists an ad for this webpage when we collect the webpage information. And we view this original ad as our groundtruth. To ensure the uniqueness of each ad in the auction, we exclude the original ad from our ad dataset.\\
%%\noindent\textbf{Step 3:} We further connect the sampled $n$ bidding prices and $n$ ads. Since the original ad will win the auction when using GSP, we assign it with the highest bidding price. Note that there exist multiple advertisers who bid with the same highest price, we make sure that the original ad will rank first when sorted by bidding price.\\
%%\noindent\textbf{Step 4:} For each ad candidate, we generate its CTR from a uniform distribution in the range of 0 and 13.97\textperthousand $+$ 1.96 $\times$ 28.49\textperthousand, which has been discussed in section \ref{sec:data_description}. 

The above five steps allow us to have enough information to compute the mentioned six metric variables for each ad in each auction. We then re-rank ads based on the trade-offs preference, which is expressed in terms of weights. In experiments, we perform the 10-fold cross validation method to obtain the optimal weights as well as to provide the performance analysis.   

% To selected a proper ad from the candidates, for each pair of webpage-ad, we parallelly perform traditional GSP auction, randomly generate CTR, predict memorability score of ad image, measure semantic matching score and calculate ad saliency. Then we fuse all the above 6 attributes using a multimodal approach as described in section \ref{sec:re-ranking}. The ad with highest ranking score will be selected as proper ad. We compared the performance of our proposed method with the traditional GSP strategy \cite{}, which selects ads with highest bidding price. To demonstrate the effectiveness of our proposed method, we try different weights of the re-ranked variables. 

% ++++++++++++++++++++++++++++++++++++++++++++++++++
\vspace{-5pt}
\subsection{Variable Analysis}
\label{sec:variable_analysis}

%We first examine two multimedia datasets and find that the growth trends of unique ads with webpages follow a format of logarithm function. 

%Fig.~\ref{fig:unique_ad_with_webpage} summarises descriptive statistics of variables across the four datasets. Although there are a large number of ads in the ad-network, only a few ads are displayed due to the SP auction. The advertisers bid for the auctions with higher prices to obtain more impressions. We find that there are some ads appearing from time to time. The repetitively displayed strategy reinforces users' memory towards the ads, but it also arises intrusiveness to users' online experience~\cite{campbell2008shut}. 

The values of metric variables in the ground truth are examined -- those ads which have been recommended to users under the SP auction model in RTB. For each impression, it contains the publisher's revenue, the advertiser's utility, the ad CTR, and the multimedia variables. Fig.~\ref{fig:variable_relationship} shows the descriptive statistics. The density plots along the diagonal show the distribution of each variable individually; the scatter plots lying in the lower triangular region show how much one variable is affected by another; and their linear correlations are given in the corresponding areas in the upper triangular part. The scatter plots lying on the bottom row shows the values of each variable under the two platforms and the right column show the variations and outliers of each variable. Note that relevance score ranges from 0 to 5, where 5 indicates most relevant and 0 indicates irrelevant, and the scores of memorability and saliency range from 0 to 1, respectively. Several findings are worth mentioning. Firstly, many displayed ads in the ground truth data are irrelevant to the contents of their hosting webpages, which will result in less user engagement. Secondly, most of displayed ads are not salient, thus can be easily overlooked. These two findings may explain the phenomenon of ad overlook in the existing advertising systems~\cite{sajjacholapunt2014influence}. We also find that pairwise variables are independent, which allows using linear combination to calculate the rank score for each ad in the auction without considering the correlation matrix.

\vspace{-10pt}
\subsection{Sensitivity Analysis}

%By introducing multimedia metric variables, together with the advertiser's utility and the ad CTR, our proposed framework evaluates the competence of advertisers' ads along different dimensions, rather than bids. 

%In our proposed two-stage framework, rather than bidding price, the rank score becomes the criteria select a proper ad from the auction. Intuitively, an ad with higher values in all the six variables will likely be selected. In this regard, our system will motivate the advertisers to better design their ads (higher values in CTR and memorability), and target the webpage and users (higher values in relevance and saliency). However, we also noticed that in some cases, an ad may never win auction eventhough we tried all the weight combinations. Alternatively, the advertiser may raise her bidding price or 
%bid for another suitable webpage.

Fig.~\ref{fig:sensitivity_analysis} presents an empirical example of analysing trade-offs among stakeholders by investigating all advertisers' ads and possible weights of metric variables. Fig.~\ref{fig:sensitivity_analysis}(a) shows the screenshot of the webpage, where ad 3010 (i.e., the ground truth) is displayed on the side bar. Since the sampled RTB auction has 9 advertisers, we further sample 8 candidate ads from the AOL dataset. They have almost the same width and length as the ground truth ad. However, they have different contents, color themes, and descriptions. Fig.~\ref{fig:sensitivity_analysis}(a)~table exhibits the computed values of variables for re-ranking, where the ground truth has been highlighted in bold. It has the highest value in revenue; and relatively high values in utility, memorability; but small values in CTR, relevance and saliency. Obviously, the ground truth is biased towards the publisher's interest. 

Radar charts shown in Fig.~\ref{fig:sensitivity_analysis}(b) provide a clearer comparison of the ground truth ad with other ad candidates. Each radar chart has multiple axes and each variable is shown as a point on the axis. A point closer to the center on an axis indicates a lower value and vice versa. In radar charts, all values of variables are further normalized into the range between 0 and 1. If all the variables' values of an ad are less than another ad, we call it being \emph{strictly dominated} because this ad will have no possibility to be selected in the second stage re-ranking. For example, ad 1319 is strictly dominated and its area in the radar chart can be fully covered by ad 3402.  

Fig.~\ref{fig:sensitivity_analysis}(c) shows all possible combinations of the weights in the re-ranking. For each subplot, x-axis represents all the weights instances that make the ad win the auction, and y-axis represents the values of weights in each instance based on stack bars. Note that the subplots of ad 1319 and ad 1847 are empty, indicating that no matter what weights we choose, these two ads will never be selected in the auction. An ad can be selected if its high valued variables have large weights. Of course, the weights represent the preference of trade-offs and different numbers of weights combination instances of ads also tell that they do not have equal chances to be selected in the re-ranking. Although ad 1847 is not strictly dominated in radar charts, it is not been selected in all weights combinations. This is because most of its variables' values are relatively small and its highest variable value is still less than that of ad 3010. Therefore, even when the trade-offs preference is biased towards the ad memorability, the rank score of ad 1847 is still less than ad 3010. We simply call it to be \emph{weakly dominated}.

\begin{figure}[t]
\small
\centering
\includegraphics[width=0.95\linewidth]{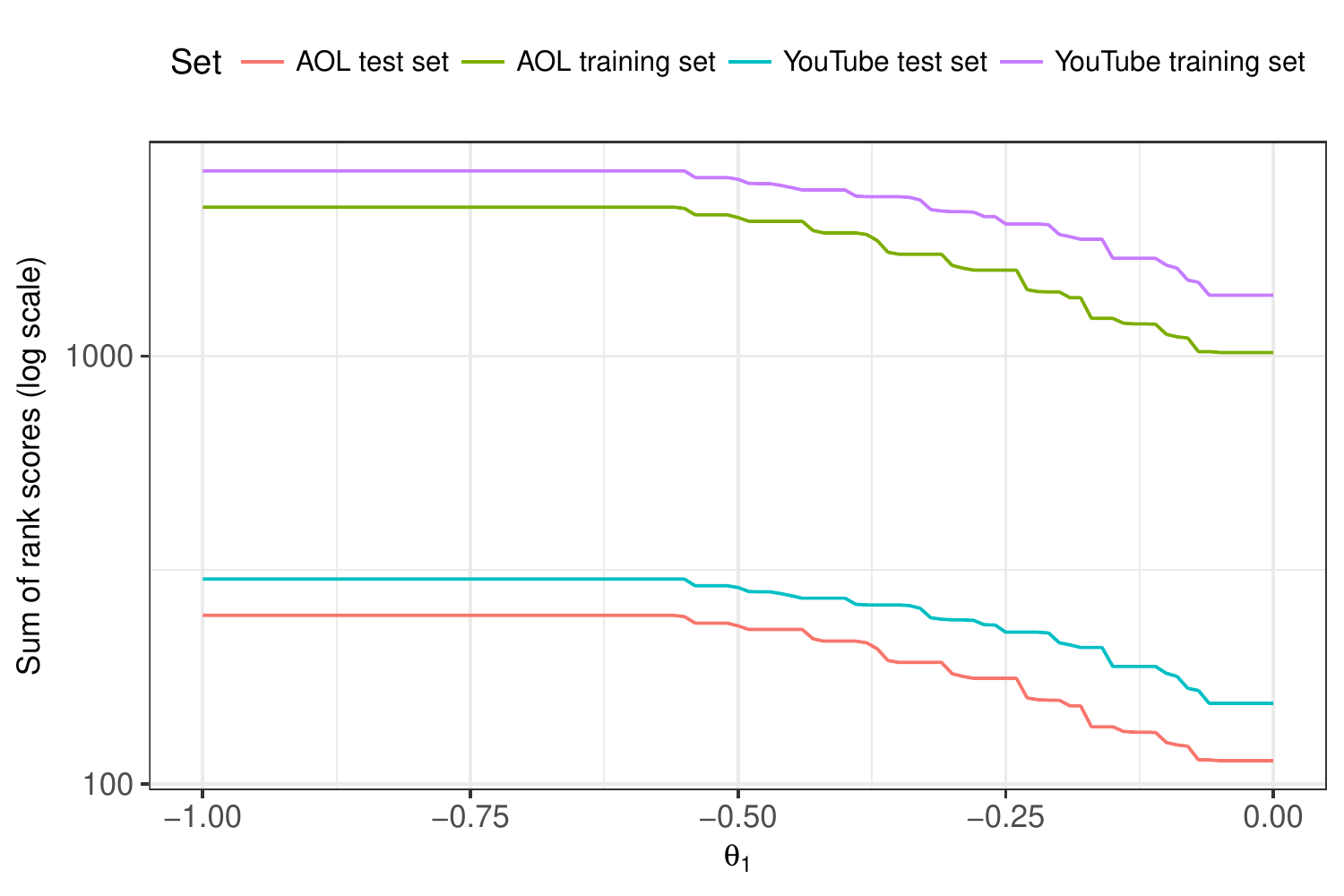}
\vspace{-7pt}
\caption{The effect of $\theta_1$ on the sum of total rank scores of the selected advertisers in auctions.}
\label{fig:theta_1_rank_score}
\vspace{10pt}
\includegraphics[width=0.95\linewidth]{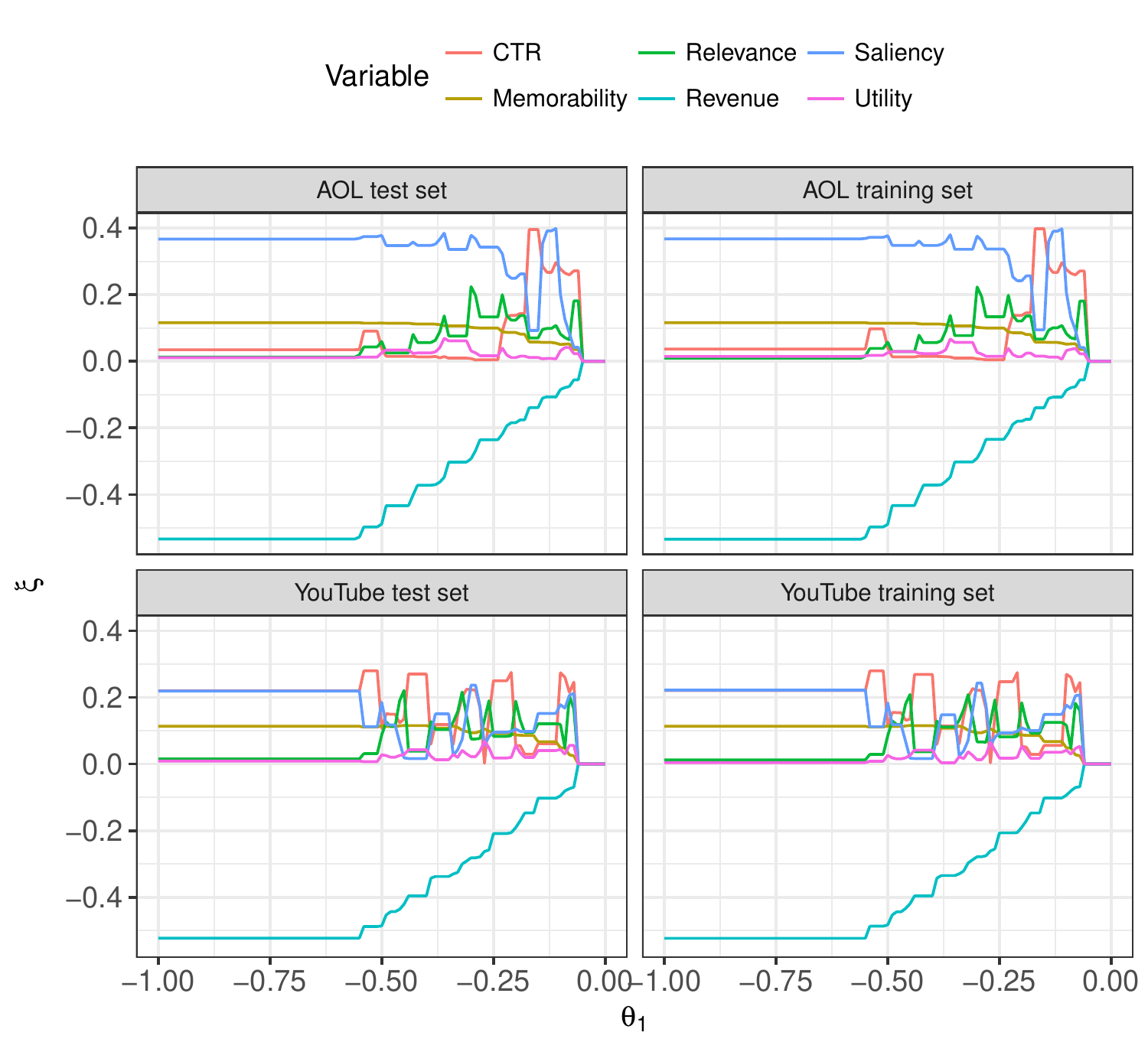}
\vspace{-7pt}
\caption{The effect of $\theta_1$ on the changes of variables.}
\label{fig:theta_1_variables}
\end{figure}

\begin{table*}[t]
\small
\centering
\caption{Summary of optimal trade-offs among stakeholders in the AOL dataset where $\theta_1 = -0.06$. Note that $\omega_{i}, i = 1,\dots,6$ represent the optimal weights obtained from training set, and $\xi_{i}, i = 1,\dots,6$ represent the changes of corresponding variables.}
\vspace{-5pt}
\label{tab:aol_overall_result}
\begin{tabular}{r|r|r|r|r|r|r|r|r|r|r|r|r|r|r|r|r|r|r}
\hline
\multirow{2}{*}{Fold}     & \multicolumn{6}{c|}{Optimal weight}	& \multicolumn{6}{c|}{Training set} & \multicolumn{6}{c}{Test set}\\
\cline{2-19}
	& $\omega_{1}$ & $\omega_{2}$ & $\omega_{3}$ & $\omega_{4}$ & $\omega_{5}$ & $\omega_{6}$ & $\xi_{1}$ & $\xi_{2}$ & $\xi_{3}$ & $\xi_{4}$ & $\xi_{5}$ & $\xi_{6}$ & $\xi_{1}$ & $\xi_{2}$ &
 $\xi_{3}$ & $\xi_{4}$ & $\xi_{5}$ & $\xi_{6}$\\
\hline
1 	& 0.45 & 0.30 & 0.05 & 0.05 & 0.15 & 0.00 & -5.5\% & 2.3\% & 3.4\% & 26.7\% & 17.4\% & 4.4\% & -5.8\% & 1.0\% & 4.0\% & 30.1\% & 23.7\% & 1.3\% \\ 
2	& 0.45 & 0.30 & 0.05 & 0.05 & 0.15 & 0.00 & -5.5\% & 2.5\% & 3.5\% & 26.7\% & 17.6\% & 4.1\% & -5.7\% & -0.7\% & 3.9\% & 30.6\%  & 22.0\% & 3.7\% \\
3	& 0.45 & 0.30 & 0.05 & 0.05 & 0.15 & 0.00 & -5.5\% & 1.8\% & 3.6\% & 27.3\% & 18.2\% & 3.7\% & -5.7\% & 6.3\% & 3.0\% & 25.3\% & 16.7\% & 7.6\% \\
4 	& 0.45 & 0.30 & 0.05 & 0.05 & 0.15 & 0.00 & -5.5\% & 1.9\% & 3.5\% & 27.5\% & 17.8\% & 4.1\% & -5.1\% & 4.8\%& 3.3\%  & 23.4\% & 19.7\% & 4.3\% \\
5	& 0.45 & 0.30 & 0.05 & 0.05 & 0.15 & 0.00 & -5.5\% & 2.5\% & 3.4\% & 27.3\% & 18.0\% & 4.1\% & -5.5\% & 0.1\% & 4.1\% & 25.4\% & 18.5\% & 3.8\% \\
6	& 0.45 & 0.30 & 0.05 & 0.05 & 0.15 & 0.00 & -5.6\% & 2.4\% & 3.6\% & 27.2\% & 18.2\% & 4.1\% & -5.0\% & 0.9\% & 2.9\% & 26.3\% & 16.2\% & 4.2\% \\
7   & 0.45 & 0.30 & 0.05 & 0.05 & 0.15 & 0.00 & -5.5\% & 2.2\% & 3.5\% & 26.8\% & 18.2\% & 4.3\% & -5.9\% & 2.5\% & 3.4\% & 29.3\% & 16.7\% & 2.4\% \\
8	& 0.45 & 0.30 & 0.05 & 0.05 & 0.15 & 0.00 & -5.6\% & 2.1\% & 3.6\% & 27.4\% & 18.4\% & 4.4\% & -4.9\% & 3.4\% & 2.6\% & 24.2\% & 14.6\% & 1.5\% \\
9	& 0.45 & 0.30 & 0.05 & 0.05 & 0.15 & 0.00 & -5.5\% & 2.2\% & 3.5\% & 27.2\% & 18.3\% & 4.0\% & -5.3\% & 2.7\% & 3.8\% & 25.7\% & 15.6\% & 4.7\% \\
10	& 0.45 & 0.30 & 0.05 & 0.05 & 0.15 & 0.00 & -5.4\% & 2.3\% & 3.5\% & 26.6\% & 18.1\% & 3.7\% & -6.1\% & 1.3\% & 4.0\% & 30.8\% & 17.0\% & 8.0\% \\
\hline
Mean	& - & - & - & - & - & - & -5.5\% & 2.2\% & 3.5\% & 27.1\% & 18.0\% & 4.1\% & -5.5\% & 2.2\% & 3.5\% & 27.1\% & 18.1\% & 4.1\% \\ 
Std.	& - & - & - & - & - & - & 0.000 & 0.002 & 0.000 & 0.002 & 0.003 & 0.002 & 0.003 & 0.020 & 0.004 & 0.026 & 0.027 & 0.021\\
\hline
\end{tabular}
\vspace{15pt}
\caption{Summary of optimal trade-offs among stakeholders in the YouTube dataset where $\theta_1 = -0.07$. Note that $\omega_{i}, i = 1,\dots,6$ represent the optimal weights obtained from training set, and $\xi_{i}, i = 1,\dots,6$ represent the changes of corresponding variables.}
\vspace{-5pt}
\label{tab:youtube_overall_result}
\begin{tabular}{r|r|r|r|r|r|r|r|r|r|r|r|r|r|r|r|r|r|r}
\hline
\multirow{2}{*}{Fold}   & \multicolumn{6}{c|}{Optimal weight}	& \multicolumn{6}{c|}{Training set} & \multicolumn{6}{c}{Test set}\\
\cline{2-19}
	& $\omega_{1}$ & $\omega_{2}$ & $\omega_{3}$ & $\omega_{4}$ & $\omega_{5}$ & $\omega_{6}$ & $\xi_{1}$ & $\xi_{2}$ & $\xi_{3}$ & $\xi_{4}$ & $\xi_{5}$ & $\xi_{6}$ & $\xi_{1}$ & $\xi_{2}$ &
 $\xi_{3}$ & $\xi_{4}$ & $\xi_{5}$ & $\xi_{6}$\\
\hline
1 	& 0.45 & 0.25 & 0.10 & 0.00 & 0.15 & 0.05 & -6.2\% & 0.3\% & 4.9\% & 0.0\% & 23.5\% & 17.9\% & -6.2\% & 0.5\% & 4.7\% & -2.7\% & 30.5\% & 18.2\% \\ 
2	& 0.45 & 0.30 & 0.05 & 0.05 & 0.10 & 0.05 & -6.9\% & 5.7\% & 2.2\% & 27.2\% & 14.7\% & 20.9\% & -6.8\% & 7.2\% & 1.3\% & 27.4\% & 18.9\% & 21.8\% \\ 
3	& 0.45 & 0.30 & 0.05 & 0.05 & 0.10 & 0.05 & -6.9\% & 5.8\% & 2.1\% & 27.4\% & 14.6\% & 21.0\% & -6.8\% & 6.8\% & 2.6\% & 25.7\% & 19.2\% & 21.1\% \\ 
4 	& 0.45 & 0.30 & 0.05 & 0.05 & 0.10 & 0.05 & -6.8\% & 5.4\% & 2.1\% & 27.0\% & 15.4\% & 21.1\% & -7.2\% & 9.5\% & 2.4\% & 29.1\% & 12.5\% & 20.1\% \\
5	& 0.45 & 0.30 & 0.05 & 0.05 & 0.10 & 0.05 & -6.8\% & 5.9\% & 2.1\% & 27.4\% & 15.4\% & 20.9\% & -7.2\% & 5.9\% & 2.3\% & 24.9\% & 12.1\% & 22.5\% \\ 
6	& 0.45 & 0.30 & 0.05 & 0.05 & 0.10 & 0.05 & -6.9\% & 6.2\% & 2.2\% & 27.3\% & 15.4\% & 20.5\% & -6.8\% & 3.4\% & 2.0\% & 25.7\% & 12.4\% & 25.8\% \\ 
7   & 0.45 & 0.30 & 0.05 & 0.05 & 0.10 & 0.05 & -6.8\% & 6.0\% & 2.1\% & 26.8\% & 15.1\% & 20.8\% & -7.2\% & 4.2\% & 2.4\% & 30.7\% & 15.1\% & 23.1\% \\ 
8	& 0.45 & 0.30 & 0.05 & 0.05 & 0.10 & 0.05 & -6.8\% & 5.9\% & 2.1\% & 27.2\% & 15.2\% & 21.1\% & -7.2\% & 5.4\% & 2.9\% & 26.6\% & 14.4\% & 20.8\% \\
9   & 0.45 & 0.30 & 0.05 & 0.05 & 0.10 & 0.05 & -6.9\% & 5.8\% & 2.2\% & 27.1\% & 14.9\% & 21.1\% & -6.9\% & 6.3\% & 1.8\% & 27.8\% & 16.5\% & 20.3\% \\ 
10  & 0.45 & 0.30 & 0.05 & 0.05 & 0.10 & 0.05 & -6.9\% & 5.8\% & 2.2\% & 26.9\% & 15.3\% & 21.4\% & -6.9\% & 6.5\% & 1.8\% & 30.3\% & 12.9\% & 17.7\% \\ 
\hline
Mean	& - & - & - & - & - & - & -6.8\% & 5.3\% & 2.4\% & 24.4\% & 16.0\% & 20.7\% & -6.9\% & 5.6\% & 2.4\% & 24.5\% & 16.5\% & 21.1\% \\ 
Std.	& - & - & - & - & - & - & 0.001 & 0.016 & 0.008 & 0.081 & 0.025 & 0.009 & 0.002 & 0.023 & 0.008 & 0.092 & 0.053 & 0.022\\
\hline
\end{tabular}
\end{table*}

% ++++++++++++++++++++++++++++++++++++++++++++++++++
\subsection{Overall Results}

We validate the proposed framework by conducting the 10-fold cross validation on both AOL and YouTube datasets. The optimal weights are estimated from the training set; they are then used for re-ranking ads in the auctions in the test set. For simplicity but without lose of generality, the threshold value $\theta_k$ is set to be zero for any $k \in K \setminus\{1\}$ so that the effects of $\theta_1$ can be examined explicitly. As described earlier, $\theta_1$ is the maximum decrease of the publisher's revenue, which indicates how much the publisher would like to sacrifice in order to increase the benefits of the other two stakeholders. Therefore, $\theta_1$ lies in the interval between -1 and 0. The smaller the value of $\theta_1$, the more is the increase in the benefits of other stakeholders. 

Fig.~\ref{fig:theta_1_rank_score} presents the effect of $\theta_1$ on the sum of rank scores of the selected advertisers in auctions. When $\theta_1$ decreases, the sum of rank scores increases. This monotone decreasing pattern is because that the solution space of weights increases if $\theta_1$ decreases (see Section~\ref{sec:optimal_weights}). Also, in the 10-fold cross validation, the size of the training set is 9 times of the test set, the solution space of weights and the sum of rank scores in the training set are roughly 9 times of the test set. The effects of $\theta_{1}$ on the changes of metric variables are examined in Fig.~\ref{fig:theta_1_variables}. In both AOL and YouTube datasets, the changes of variables in the training data follow the same patterns as those in the test data. If $\theta_{1}$ is close to $0$, the SP auction (the ground truth) is the optimal solution. This means that our model suggests that the publisher shouldn't sacrifice any revenue. If the publisher further decreases $\theta_{1}$, the optimal weights change. There will be some growth in the variables of other stakeholders, and of course, the publisher's revenue starts to decrease. If $\theta_{1}$ is small, the publisher's revenue will decrease significantly and she may find it unacceptable. As shown in Fig.~\ref{fig:theta_1_variables}, if $\theta_{1}$ is less than -0.3, it reduces the publisher's revenue by almost 30\%. There is no monotone increasing pattern in other variables. Analyzing the changes in variables will help the publisher to decide a proper $\theta_{1}$. Tables~\ref{tab:aol_overall_result}-\ref{tab:youtube_overall_result} further present two special cases: $\theta_{1} = -0.06$ for the AOL dataset; and $\theta_{1} = -0.07$ for the YouTube dataset. Both cases show the results when the decrease in the publisher's revenue is close to the pre-specified threshold value $\theta_1$. For example, there is about 5.5\% decrease in the publisher's revenue in the AOL training set and other variables enjoy certain increases ranging from 2.2\% to 27.1\%.

% ++++++++++++++++++++++++++++++++++++++++++++++++++
% +++++++++++++++++++++++++++++++++++++++++++++++++
\section{Conclusion}
\label{sec:conclusion}

This paper discusses a two-stage framework that optimizes trade-offs among stakeholders in display advertising. Different to many related studies which only focus on the publisher's revenue, we further consider the benefits of the advertiser and the user by incorporating multimedia metrics. The trade-offs optimization is based on a linear combination of all metric variables and their weights are learnt from data. Variable and sensitivity analysis are conducted to explore the solution space. Our experimental results validate the proposed framework and show that it is able to increase the benefits of other two stakeholders with just a slight decrease in the publisher's revenue. In the long run, better engagement of advertisers and users will increase the demand of advertising and supply of webpage visits, which can then boost the publisher's revenue. Future research can potentially discuss a framework which considers a generalized scenario where a webpage has multiple ad slots being separately auctioned off in RTB. The long-term effects of trade-offs optimization can also be further investigated and it would be very interesting if the framework can be tested over a live setting.

% ++++++++++++++++++++++++++++++++++++++++++++++++++
% ++++++++++++++++++++++++++++++++++++++++++++++++++
\section{Acknowledgement}
This research is supported by the National Research Foundation, Prime Minister's Office, Singapore under its International Research Centre in Singapore Funding Initiative.

% ++++++++++++++++++++++++++++++++++++++++++++++++++
% ++++++++++++++++++++++++++++++++++++++++++++++++++
\bibliographystyle{abbrv}
\bibliography{mybib}

% ++++++++++++++++++++++++++++++++++++++++++++++++++
% ++++++++++++++++++++++++++++++++++++++++++++++++++
\end{document}